\documentclass[%
 reprint,
 amsmath,amssymb,
 aps,
 prl,
]{revtex4-2}

\usepackage{graphicx}% Include figure files
\usepackage{dcolumn}% Align table columns on decimal point
\usepackage{bm}
\usepackage{xcolor}
\begin{document}

\preprint{APS/123-QED}

\title{Coupled instabilities drive quasiperiodic order-disorder transitions in Faraday waves}

\author{Valeri Frumkin}
\email{valerafr@mit.edu}
\thanks {These authors contributed equally to this work}
\affiliation{Department of Mathematics, Massachusetts Institute of Technology}

\author{Shreyas Gokhale}
\email{gokhales@mit.edu}
\thanks {These authors contributed equally to this work}
\affiliation{Department of Physics, Massachusetts Institute of Technology}

%\author{John W. M. Bush}
% \email{bush@math.mit.edu}
%\affiliation{Department of Mathematics, Massachusetts Institute of Technology.}

\date{}

\begin{abstract}
We present an experimental study of quasiperiodic transitions between a highly ordered square-lattice pattern and a disordered, defect-riddled state, in a circular Faraday system. We show that the transition is driven initially by a long-wave amplitude modulation instability, which excites the oscillatory transition phase instability, leading to the formation of dislocations in the Faraday lattice. The appearance of dislocations damps amplitude modulations, which prevents further defects from being created and allows the system to relax back to its ordered state. The process then repeats itself in a quasiperiodic manner. Our experiments reveal a surprising coupling between two distinct instabilities in the Faraday system, and suggest that such coupling may provide a generic mechanism for quasiperiodicity in nonlinear driven dissipative systems.   
\end{abstract}

\maketitle

When a thin layer of fluid is subjected to uniform vertical vibration with sufficiently large amplitude, the initially flat fluid surface destabilizes to an ordered pattern of sub-harmonic standing waves, known as Faraday waves \cite{faraday_xvii_1831}. The Faraday system has been the subject of numerous theoretical \cite{benjamin_stability_1954,miles_parametrically_1990,rajchenbach_faraday_2015} and experimental \cite{edwards_parametrically_1993,arbell_pattern_2002, westra_patterns_2003} studies, and it serves as a canonical example of a nonlinear pattern-forming system \cite{cross_pattern_1993,gollub_pattern_1999}. Its importance, however, goes beyond the study of pattern formation, as it manifests in a wide range of physical systems across multiple length scales. Faraday waves have been observed in systems as disparate as Bose-Einstein condensates \cite{engels_observation_2007}, soft elastic solids \cite{bevilacqua_faraday_2020}, and even  bodies of vibrated living earthworms \cite{maksymov_excitation_2020}. In pilot-wave hydrodynamics, locally excited Faraday waves store information about the trajectories of walking droplets \cite{eddi_information_2011,bush_pilot-wave_2015,bush_hydrodynamic_2020}, while in hydrodynamic superradiance they serve as the underlying mechanism for sinusoidal oscillations of the droplet emission rate \cite{frumkin_superradiant_2022}. 

Since the Faraday system is readily accessible in the lab, it allows for a detailed study of the complex transition from order to disorder in pattern-forming systems. Specifically, when the driving amplitude is increased well beyond the Faraday threshold, defects appear in the ordered Faraday lattice, leading to the emergence of spatial disorder through a process that came to be known as ``defect-mediated turbulence" \cite{coullet_form_1989}. 
Defect formation typically occurs via secondary instabilities, such as transverse amplitude modulation (TAM) instability \cite{Ezerskii_Spatiotemporal_1986,tufillaro_order-disorder_1989,milner_square_1991}, and the oscillatory transition phase (OTP) instability \cite{shani_localized_2010}. In the former, the square Faraday pattern is modulated by long wavelength oscillations normal to the air-fluid interface, leading to an eventual loss of long-range order with increasing driving amplitude. In the latter, spatially uncorrelated elastic waves are excited within the plane of the Faraday lattice, leading to the emergence of defects. In both cases, as the defects are formed, the square Faraday pattern exhibits a state of spatial intermittency where the ordered and disordered phases can coexist. With further increase in driving amplitude the pattern loses any long range order and ``melts" into a fully spatiotemporally disordered state \cite{bosch_spatiotemporal_1993,goldman_lattice_2003}.

A less known, but intriguing phenomenon is that of temporal intermittency in the order-disorder transition in the Faraday system. This phenomenon was first reported by Ezerskii \cite{ezersky_temporal_1991}, who observed that when the depth of the liquid was chosen such that the group velocity of capillary waves was close to the velocity of low-frequency gravity modes, $C=\sqrt{\rho g}$, resonant conditions would occur allowing for efficient energy transfer between the two. As a result, in a specific parameter regime, weakly damped gravity modes would get excited and slowly grow in amplitude, leading to an accelerated generation of higher harmonics and a rapid transition to chaos. The system would then alternate quasi-periodically between the low frequency oscillations and fully disordered high frequency modes. This behaviour was independent of the system's geometry, and the only condition for its emergence was the aforementioned resonance between capillary waves and the weakly damped gravity modes. 

Here we describe a different path to quasiperiodic dynamics in the Faraday system. We show that in the case of a circular bath, for specific values of the bath radius, amplitude modulations in the shape of vibrational modes of a circular elastic membrane are excited in the Faraday lattice. These modes are resonantly amplified by the driving, leading to the formation of dislocations via the OTP instability. The increased dislocation density leads to a rapid decay of spatial correlations, preventing formation of any additional dislocations, and allowing the system to relax back to its ordered state. The process then repeats itself. The phenomenon described here reveals a surprising coupling between two distinct instabilities in the Faraday system, namely, TAM and OTP. 

\begin{figure*}
  \centering
  \includegraphics[width=1 \textwidth]{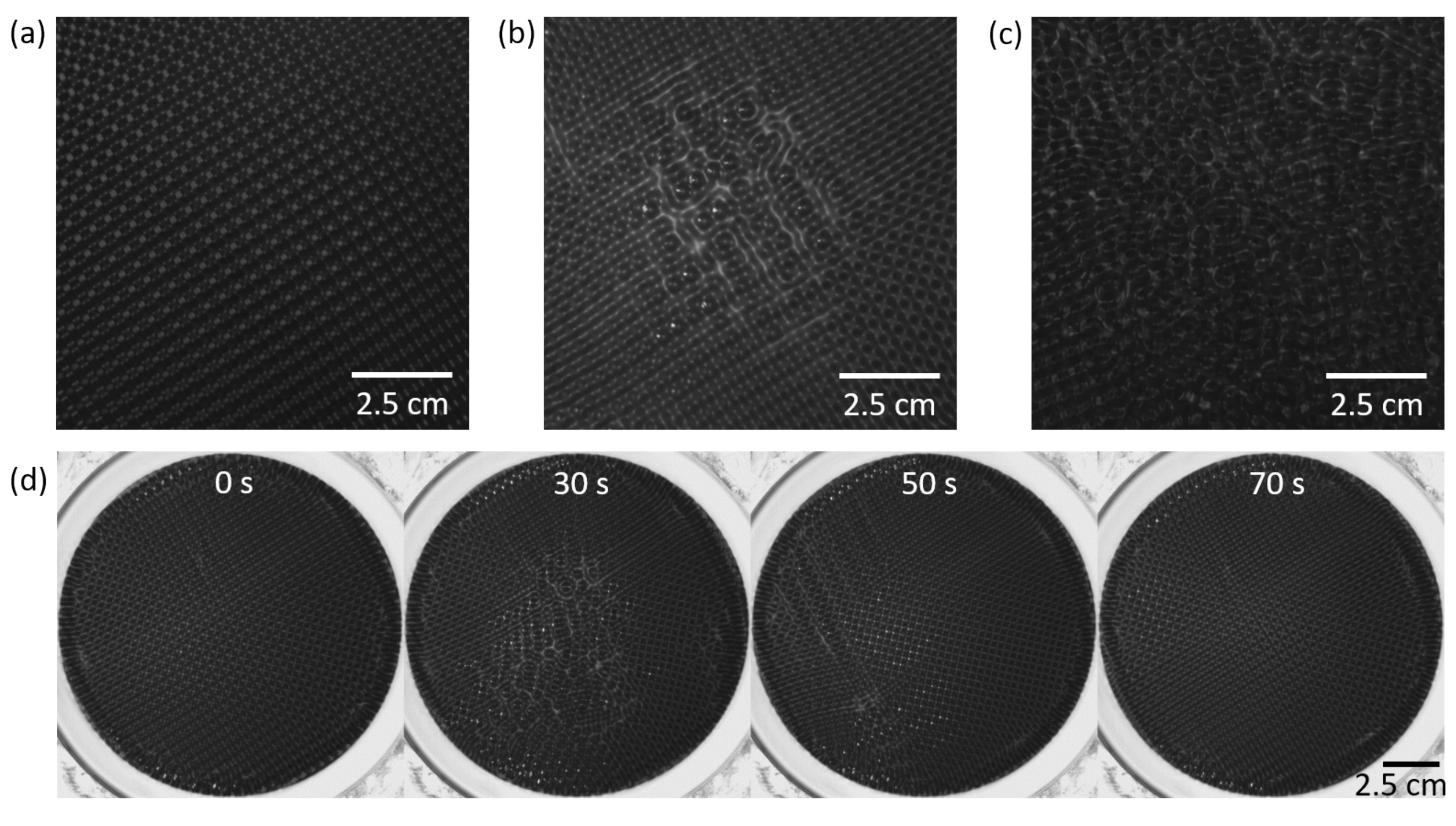}
  \caption{{\bf Quasiperiodic transitions between order and disorder in Faraday waves.} (a-c) Snapshots of our system for a fixed driving frequency $f_d = 88$ Hz showing an ordered Faraday wave lattice at peak vibration acceleration $\gamma = 5.4g$ (a), an intermittent partially disordered state at $\gamma = 5.95g$ (b), and a chaotic state at $\gamma = 6.5g$. (d) A time sequence of snapshots during a typical quasiperiod from the same data set as in (b), showing the onset of disorder followed by the clearing of defects and reordering of the lattice.}
  \label{fig1}
\end{figure*}

Our experimental system consisted of a circular bath, $190$ mm in diameter, that contained a $5$ mm deep circular opening, with a diameter of $156$ mm.
The bath was filled with silicon oil so that the resulting oil depth was $5.6 \pm 0.2$ mm above the inner opening, and $ \approx 0.6$ mm in the surrounding shallow layer.  The shallow layer acted as a wave damper and eliminated any effects due to sloshing of oil against the boundaries of the system. The silicon oil had surface tension $\sigma=0.0209$ N/m, viscosity $\nu=20$ cSt, and density $\rho=0.965\times 10^{3}\,\,\,\text{kg}/\text{m}^{3}$.
The bath was vibrated vertically by an electromagnetic shaker with forcing $F(t)=\gamma\cos(2\pi f_{d}t)$, with $f_d$ and $\gamma$ being the frequency and peak vibrational acceleration, respectively. 
We ensured spatial uniformity of the bath vibration by connecting the shaker to the bath via a steel rod coupled to a linear air bearing~\citep{harris_generating_2015}. We monitored the vibrational forcing using two accelerometers that were placed on opposite sides of the bath, ensuring a constant vibrational acceleration amplitude to within $\pm 0.002$ g.
To image the emergent surface-wave pattern, we used a semi-reflective mirror that was positioned at 45° between the bath and a charge-coupled device (CCD) camera that was mounted directly above the setup. The bath was illuminated with a diffuse-light lamp facing the mirror horizontally, yielding images with bright regions corresponding to horizontal parts of the surface, specifically extrema or saddle points. Before each experiment, we waited for 20 minutes for the system to stabilize and captured videos for at least 10 mins, at a frame rate of $f_d/4$ fps.

\begin{figure}
\hspace{-0.22 cm}
  \includegraphics[width=0.49 \textwidth]{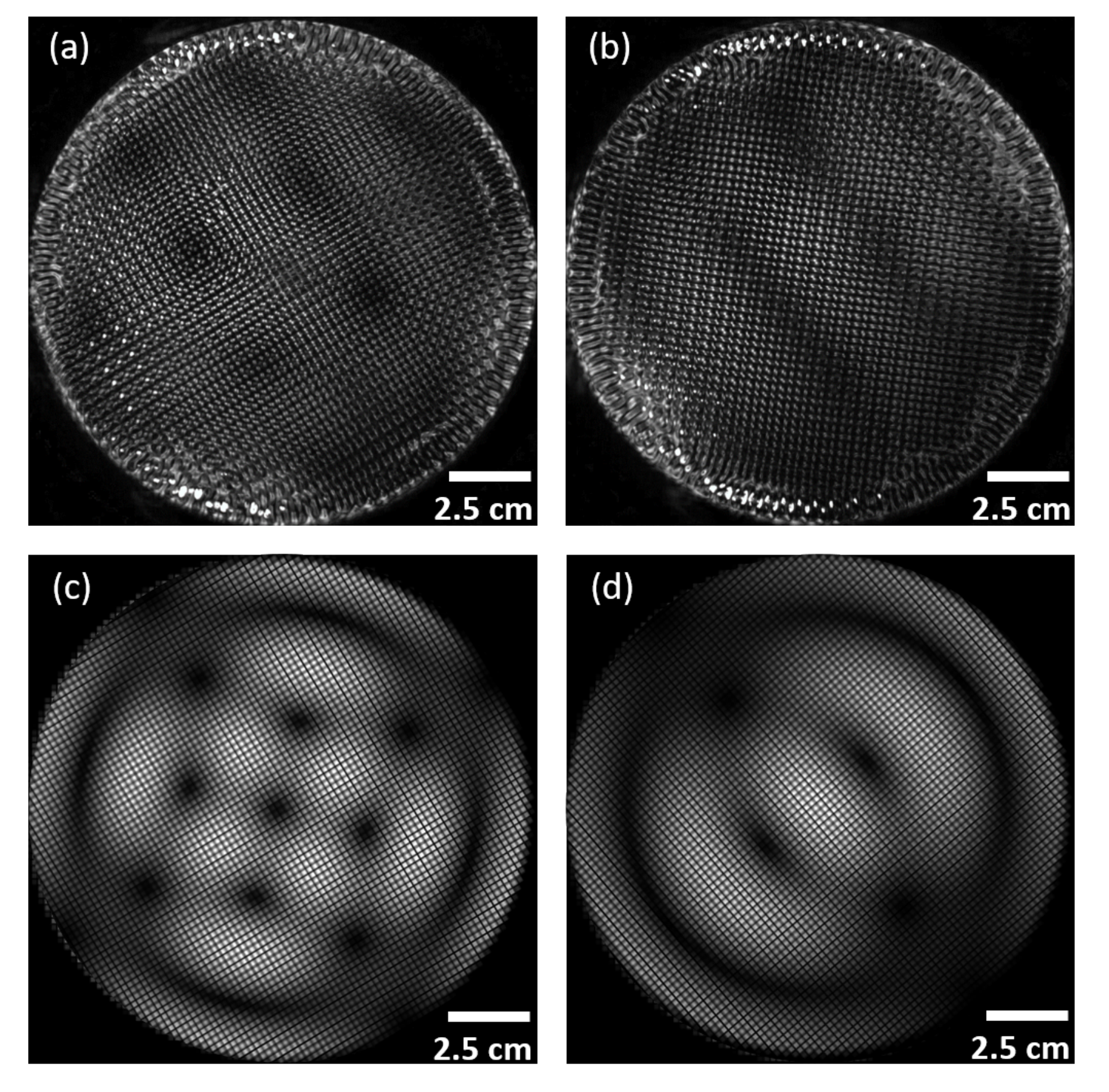}
  \caption{{\bf Spatial structure of long wavelength modes.} (a-b) Visualization of the spatial structure of long wavelength modes for $f_d = 88$ Hz (a), and $f_d = 79$ Hz (b). The images in (a) and (b) represent the pixel-wise standard deviation of image intensity over one oscillation time period ($\approx 1$ s) leading up to disordering of the lattice. The image intensity has been uniformly multiplied by $3$ to facilitate clearer visualization. (c-d) Visualization of the geometry of theoretically predicted normal modes of a circular elastic membrane, with mode numbers $(2,2)$ (c), and $(1,2)$ (d). The membrane diameter in (c-d) is chosen to be the same as for experiments shown (a-b).}
  \label{fig2}
\end{figure}

Fig.\ref{fig1} describes the typical evolution of the Faraday system with increasing driving amplitude $\gamma$. For $\gamma$ slightly above the pattern-forming threshold, the Faraday pattern takes the form of a square lattice characterized by highly coherent long-range order (Fig.\ref{fig1}a). With further increase in the driving amplitude, line dislocations appear in the lattice leading to a regime of coexistence between ordered and disordered regions, reminiscent of the intermittency route to chaos (Fig. \ref{fig1}b). The dislocation density increases with an increase in the driving amplitude, until finally the lattice ``melts" into a fully disordered, chaotic state (Fig. \ref{fig1}c). 

The behavior of the system studied here is in stark contrast with the typical intermittency route to chaos. Specifically, there appears to be a small parameter range $\gamma_1<\gamma<\gamma_2$, with $\gamma_1$ above the pattern-forming threshold and $\gamma_2$ below the dislocation-forming threshold, where amplitude modulations in the form of low frequency gravity waves are excited. These secondary waves resonate with the driving frequency and grow in amplitude overtime, leading to the formation of dislocations in the Faraday lattice. As the number of dislocations increases, the system loses its coherence and the long wavelength gravity waves are damped. In the absence of gravity waves, the formed dislocations exit the system through its boundaries, restoring the original form of a highly ordered square lattice. At this point, full coherence is restored, and the process repeats itself (Fig. \ref{fig1}d).  Over long periods of time, our system switches quasiperiodically between highly coherent ordered states, and those that are partially disordered (See Video S1). 

Notably, the low frequency waves observed here seem to represent vibrational modes of a circular membrane, which is consistent with previous observations that the ordered Faraday lattice can exhibit elastic-like behavior \cite{domino_faraday_2016}. Figure \ref{fig2}a,b shows a comparison between two clear, but distinct long wavelength modes observed at $f_d = 88$ Hz and $f_d = 79$ Hz, respectively. A comparison to theoretically predicted circular membrane modes is made in Fig. \ref{fig2}c,d, showing the $(1,2)$ and the $(2,2)$ modes, respectively. We observe a quantitative agreement between the shape and wavelength of the predicted and measured modes, further supporting this conclusion (see SI text for additional details). Modes that are excited at other frequencies in the proximity of $f_d = 88$ Hz and $f_d = 79$ Hz, exhibit some combination of the two main modes observed here (See Video S2 for a visualization of the development of these low frequency modes in our experiments). 

\begin{figure}
  \centering
  \includegraphics[width=0.5 \textwidth]{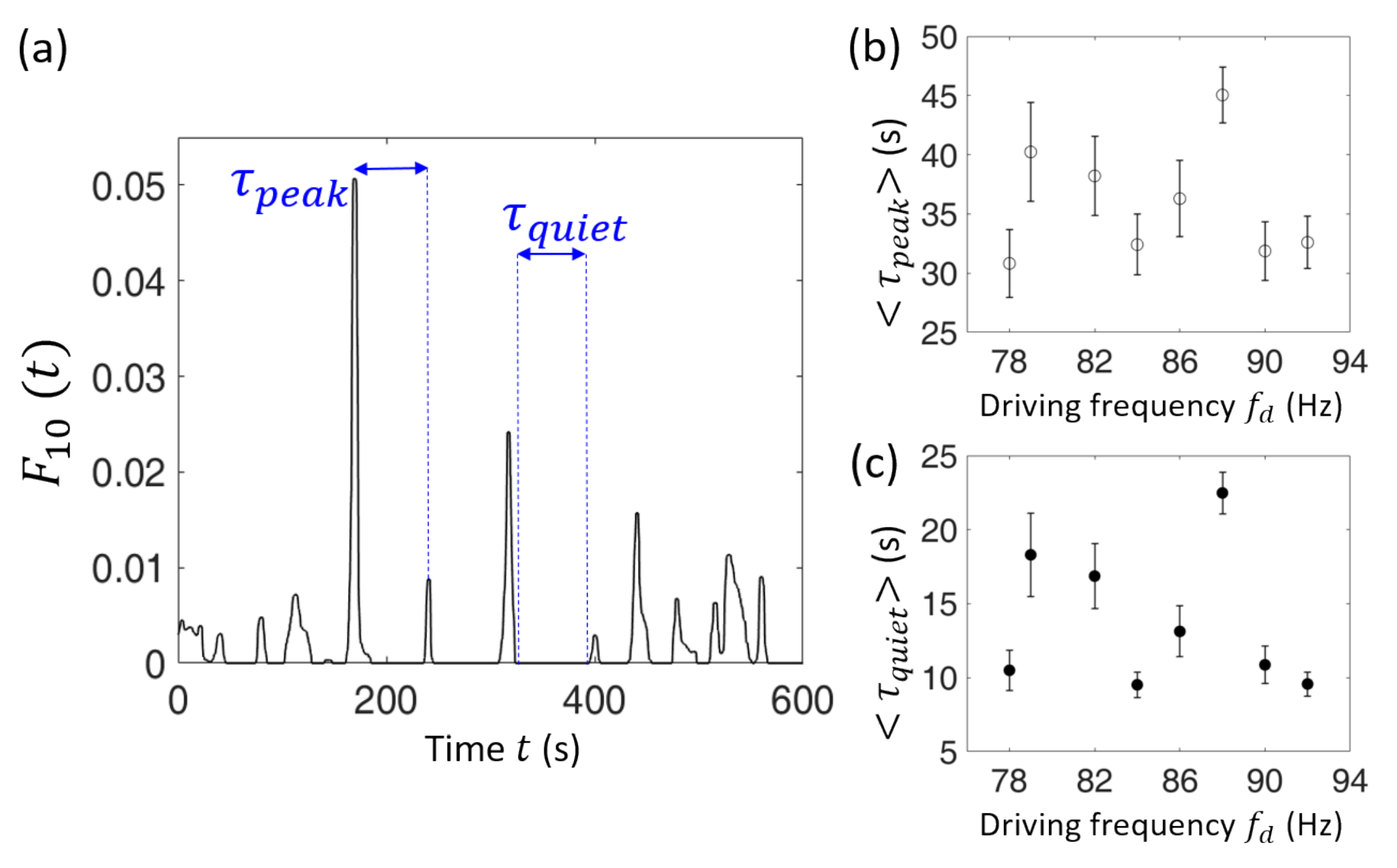}
  \caption{{\bf Quantification of timescales associated with quasiperiodic behavior.} (a) A representative time series of $F_{10}$ for $f_d = 88$ Hz in the quasiperioidic regime, showing the first $600$ s of a video of total duration $1800$ s. $F_{10}(t)$ is smoothed in time using a moving mean filter with a window of 10 frames ($0.5$ s), followed by the application of two successive moving median filters of window width 100 frames ($5$ s). (b-c) The average distance between successive peaks in $F_{10}(t)$, $\langle \tau_{\textrm{peak}} \rangle$ (b), and the average duration over which the system remains completely ordered, $\langle \tau_{\textrm{quiet}} \rangle$ (c), as a function of driving frequency $f_d$. In (b-c), error bars are standard errors of the mean of the distributions.}
  \label{fig3}
\end{figure}

In order to extract the timescales associated with quasiperiodic dynamics, we first quantified the fraction $F_{10}$ of all bright pixels that are associated with bright regions of size $> 10$ pixels (See SI text for details). As bright regions are highly localized around lattice points in ordered regions, but substantially delocalized in disordered ones, $F_{10}=0$ when the system is fully ordered, but is substantially larger when disorder is present.  A representative segment of the time series for $F_{10}$ is shown in Fig. \ref{fig3}a, for $f_d = 88$ Hz. From such time series, we extract two relevant timescales. The first, $\tau_{\textrm{peak}}$, measures the duration between successive maxima in $F_{10}$, and physically corresponds to the time between two successive bursts of disorder. The second timescale $\tau_{\textrm{quiet}}$, measures the duration over which $F_{10}$ is identically zero, and corresponds to the time over which the system persists in the fully ordered state prior to exhibiting bursts of disorder. When averaged over the entire video, the timescales $\langle \tau_{\textrm{peak}} \rangle$ (Fig. \ref{fig3}b) and $\langle \tau_{\textrm{quiet}} \rangle$ (Fig. \ref{fig3}c) both exhibit a pronounced maximum at $f_d = 88$ Hz, as well as a second maximum at $f_d = 79$ Hz. These observations are qualitatively robust to variations in smoothing parameters (Fig. S1). The maxima in $\langle \tau_{\textrm{peak}} \rangle$ and $\langle \tau_{\textrm{quiet}} \rangle$ (Fig. \ref{fig3}b-c) can be qualitatively explained by the resonant amplification of specific modes at $f_d = 88$ Hz and $f_d = 79$ Hz. At these frequencies, energy is selectively injected into resonant circular membrane modes, which suppresses disordering of the lattice due to localized short wavelength excitations. This can enable the system to remain within the ordered state over longer times, resulting in a larger $\langle \tau_{\textrm{quiet}} \rangle$ (Fig. \ref{fig3}c). The suppression of localized modes also reduces the probability of secondary dislocation bursts within the same quasiperiod, resulting in fewer peaks in $F_{10}(t)$, and therefore larger $\langle \tau_{\textrm{peak}} \rangle$ (Fig. \ref{fig3}b).

\begin{figure}
  \centering
  \includegraphics[width=0.5 \textwidth]{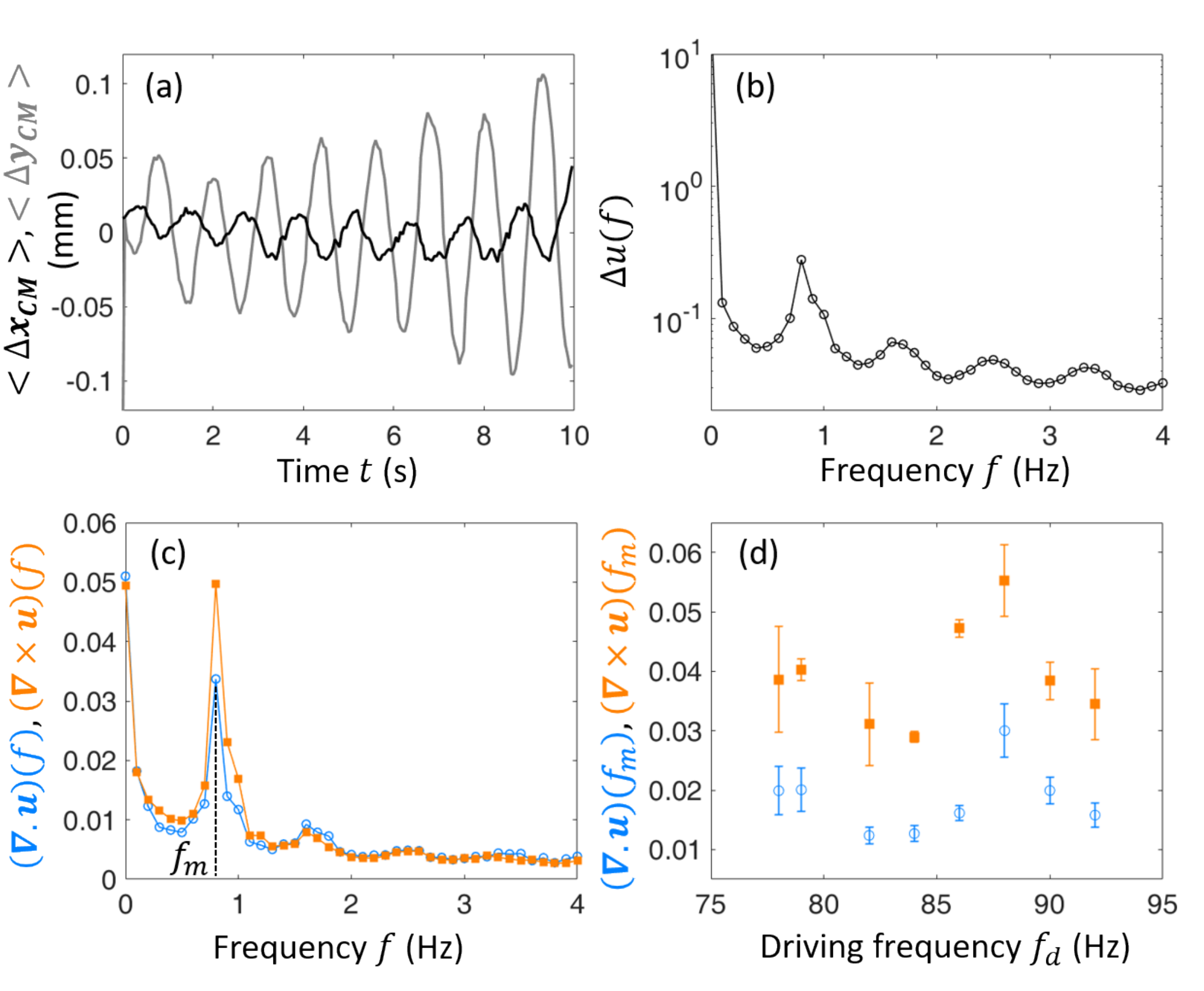}
  \caption{{\bf Enhancement of transverse fluctuations due to resonant amplification of membrane modes.} (a) Time series of the X (black) and Y (grey) coordinates of the center of mass prior to disordering of the lattice, for $f_d = 88$ Hz. (b) $\Delta u(f)$ versus frequency $f$ for the same data as shown in (a), showing a prominent peak at the oscillation frequency of the long wavelength mode.(c) Curl $(\nabla \times \bf{u})(f)$ (filled orange squares) and divergence $(\nabla \cdot \bf{u})(f)$ (open blue circles) of displacements of intensity maxima as a function of frequency $f$. The frequency corresponding to the peak, $f_m$, corresponds to the frequency of long wavelength oscillations shown in (b). (d) The peak values of curl ($(\nabla \times \bf{u})(f_m)$, filled orange squares) and divergence ($(\nabla \cdot \bf{u})(f_m)$, open blue circles) of displacements of intensity maxima as a function of driving frequency $f_d$. Error bars are standard errors of the mean across three distinct quasiperiods.}
  \label{fig4}
\end{figure}

To quantitatively characterize the oscillations leading up to disorder, we tracked the positions of intensity maxima over a period of $10$ s prior to disordering of the lattice, using Blair and Dufresne's MATLAB implementation of the Crocker-Grier algorithm \cite{crocker1996methods}. The low frequency vibrations associated with $f_d = 88$ Hz are clearly visible in the displacement of the X and Y components of the center of mass, $\langle \Delta x_{CM} \rangle$ (black) and $\langle \Delta y_{CM} \rangle$ (grey) (Fig. \ref{fig4}a). To characterize the associated lattice vibrations, following \cite{shani_localized_2010}, we quantified the Fourier spectrum of relative displacements between nearest neighbor intensity maxima. Specifically, we computed
\begin{equation}
\Delta u(f) = \sqrt{\langle \Delta x_{m,n}(f) \rangle_{m,n}^2 + \langle \Delta y_{m,n}(f) \rangle_{m,n}^2}
\end{equation}
where $\Delta x_{m,n}(f)$ and $\Delta y_{m,n}(f)$ are magnitudes of the Fourier transforms of X and Y components of relative displacements between nearest neighbors $m$ and $n$, respectively, and $\langle \rangle_{m,n}$ denotes averaging over all pairs of nearest neighbors. $\Delta u(f)$ exhibits a sharp maximum at $f \approx 0.8$ Hz (Fig. \ref{fig4}b), consistent with the low frequency vibration shown in Fig. \ref{fig4}a. A similar maximum is observed near $f \sim 1$ Hz for all driving frequencies studies (Fig. S2). 
Crucially, the sharp maximum in $\Delta u(f)$ is quite distinct from the appearance of a broad shoulder observed by Fineberg and coworkers \cite{shani_localized_2010} in the oscillatory transition phase. To further compare and contrast our results with prior work, we computed the magnitude of the Fourier transform of the curl and divergence of displacements $\bf{u}(\bf{r},t)$, $(\nabla \times \bf{u})(f)$ and $(\nabla \cdot \bf{u})(f)$, respectively, averaged over space. In the oscillatory transition phase \cite{shani_localized_2010}, no enhancement is observed for $(\nabla \cdot \bf{u})(f)$, whereas a broad spectrum of frequencies are excited for $(\nabla \times \bf{u})(f)$. This corresponds to damped oscillatory waves with purely transverse polarization. In stark contrast, we observe that both $(\nabla \cdot \bf{u})(f)$ and $(\nabla \times \bf{u})(f)$ develop a sharp peak at a characteristic frequency (Fig. \ref{fig4}c). Furthermore, we consistently observe that the peak in curl has a larger amplitude than the peak in divergence for all driving frequencies studied (Fig. \ref{fig4}d). 

In practice, the signal in $(\nabla \cdot \bf{u})(f)$ derives from low frequency modulations in the height of the fluid-air interface, which lead to apparent in-plane contractions and dilations in the positions of intensity maxima. However, the fact that these modulations lead to an enhancement in $(\nabla \times \bf{u})(f)$ suggests that low frequency gravity wave modulations lead to resonant amplification of certain modes associated with the transverse oscillatory instability observed in \cite{shani_localized_2010}. Importantly, $(\nabla \cdot \bf{u})(f)$ as well as $(\nabla \times \bf{u})(f)$ exhibit a pronounced maximum at $f = 88$ Hz, and a weaker one at $f = 79$ Hz, strongly resembling the peaks observed in Fig. \ref{fig3}b-c.  

Once in-plane oscillations are sufficiently strongly excited, they lead to increased distortions of the Faraday wave lattice. The instability continues to grow until the ordered lattice can no longer accommodate the strain due to distortions, and we observe a sudden burst of dislocations. The disordering of the lattice leads to decoherence, and hence damping, of the circular membrane mode as well as transverse oscillations. Eventually, dislocations migrate to the boundaries of the bath, allowing the lattice to be fully ordered again, and the process repeats itself. We quantitatively characterized the sequence of increasing and decreasing coherence for the order-disorder-order cycle shown in Fig. \ref{fig1}d using spatial velocity correlations. 

\begin{figure}
\hspace{-0.4 cm}
  \includegraphics[width=0.5 \textwidth]{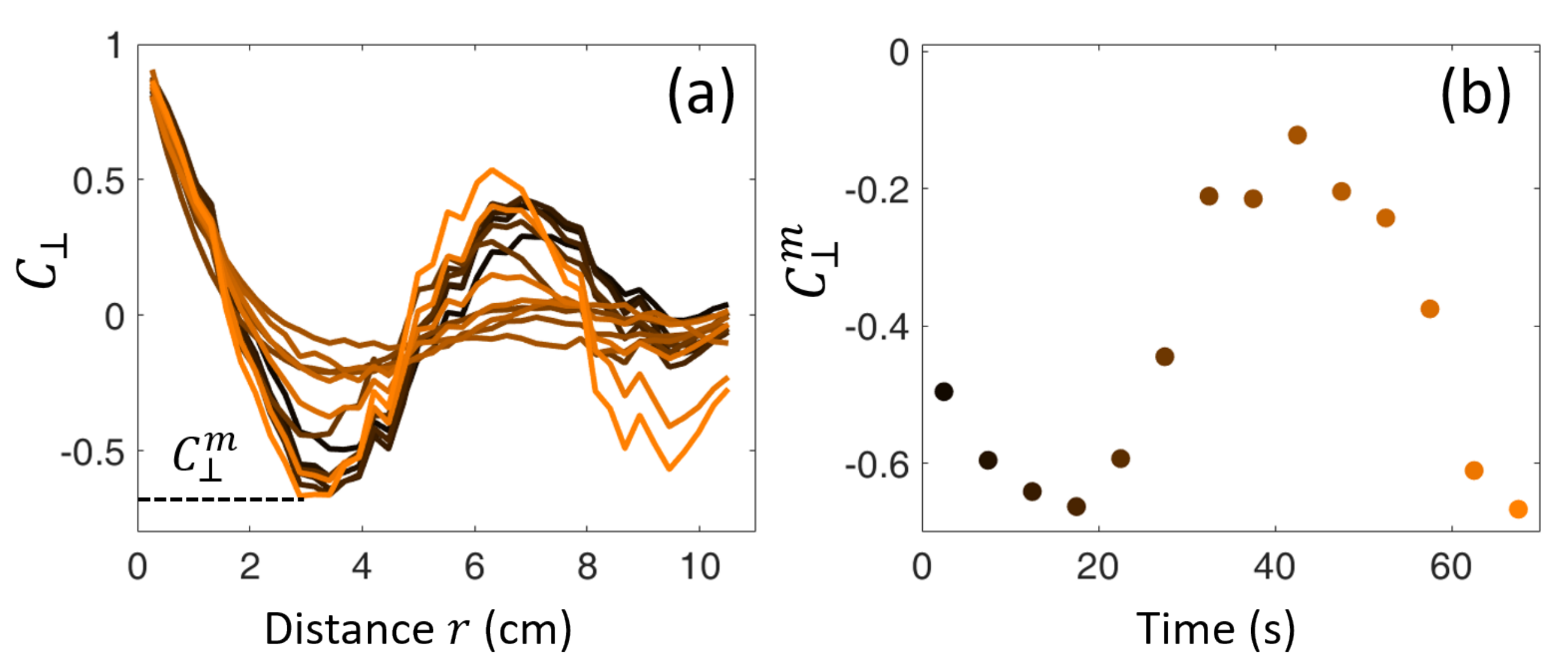}
  \caption{{\bf Increase and decrease of coherence of long wavelength modes within a quasiperiod.} (a) The equal time transverse spatial velocity correlation $C_{\perp}$ averaged over time intervals of duration $5$ s (100 frames) for the $f_d = 88$ Hz data shown in Fig. \ref{fig1}d. The color changes with time from $0$ s - $5$ s (black) to $65$ s - $70$ s (light orange). The strongly negative correlations are consistent with the predominantly transverse polarization of long wavelength oscillations. The more negative the minimum value $C_{\perp}^{m}$, the stronger are the correlations. (b) $C_{\perp}^{m}$ as a function of time during the quasiperiod. Colors correspond to the data in (a).}
  \label{fig5}
\end{figure}

We generated the velocity field for intensity maxima using the MATLAB package PIVlab \cite{thielicke2014flapping,thielicke2014pivlab,thielicke2021particle}. We then split our video into segments of duration $5$ s (100 frames), and computed the equal time transverse spatial velocity correlation $C_{\perp} = \langle v_{\perp}(0) v_{\perp}(r) \rangle$ for each segment. Here, $v_{\perp}$ denotes the component of velocity perpendicular to the line joining two points on the PIV grid, and $\langle \rangle$ denotes averaging over the time interval of $5$ s. Our choice of transverse velocity correlations was motivated by the fact that lattice vibrations have a predominantly transverse character (Fig. \ref{fig4}c,d). Fig. \ref{fig5}a shows the transverse velocity correlation for all $5$ s time intervals within the $70$ s duration shown in Fig. \ref{fig1}d. At early (black) as well as late (light orange) times, when the lattice is ordered, we observe strong correlations that span almost the entire system. At intermediate times (brown), however, the strength of the correlations is reduced significantly due to disorder. To quantify the evolution of spatial coherence, we plotted the minimum value $C_{\perp}^{m}$ of the transverse spatial velocity correlations as a function of time (Fig. \ref{fig5}b). $C_{\perp}^{m}$ initially becomes increasingly strongly negative, indicating increasingly strong correlations, as the membrane mode vibrations amplify transverse oscillations. Once the lattice disorders, $C_{\perp}^{m}$ quickly rises towards $0$, as the oscillations lose coherence. At late times, the system becomes ordered again, and resonant amplification of the membrane mode once again results in $C_{\perp}^{m}$ reaching a strongly negative value.

We have described how a quasiperiodic order-disorder transition occurs in the Faraday system as a result of coupling between two distinct instabilities. The first instability appears in the form of transverse amplitude modulations that ``feeds" the oscillatory transition phase instability through which defects are generated in the lattice. The increase in defect density damps amplitude modulations and ``starves" the OTP instability, allowing the system to restore itself back to a fully ordered state.   
Since both instabilities were previously shown to be quite general in systems that exhibit defect mediated turbulence, we expect similar quasiperiodic behaviour to also be of a general nature. 

It is interesting to note two key differences between the system considered here, and the original observations made by Ezerskii \cite{ezersky_temporal_1991}. In our experiments, global membrane modes were only excited for specific values of the bath's radius. Decreasing the radius by $5$ mm while keeping the same depth and driving frequency, did not produce similar oscillations. This is in contrast with Ezerskii's observations that the phenomenon followed purely from the resonance conditions between the weakly damped gravity modes and capillary waves, and thus did not depend on the system's boundaries. In addition, in Ezerskii's experiments, disorder appeared via higher harmonics in the TAM instability, while in our system disorder enters through local defects generated by the OTP instability. The incompatibility of these two aspects in the two experiments, suggest that they represent two distinct, although clearly related phenomena. 

Another interesting feature of our system is that the driving frequency ($79$ Hz $\leq f_d \leq 92$ Hz) is orders of magnitude higher than the observed frequency of quasiperiods ($\sim 0.02$ Hz). The characteristic timescales that give rise to quasiperiodicity in our system are therefore entirely emergent in nature, and derive solely from the positive and negative feedback between secondary instabilities. It is worth investigating whether, and to what extent, the mechanism for quasiperiodicity observed here applies to a broader class of driven-dissipative systems.

Finally, we note that our results bear intriguing connections to observation of quasiperiodic strain bursts \cite{papanikolaou2012quasi,cui2016controlling} associated with dislocation avalanches \cite{papanikolaou2017avalanches,csikor2007dislocation} in crystal plasticity. In crystal plasticity, the nonequilibrium drive is typically provided by an external mechanical deformation such as shear. Quasiperiodicity results from the interplay between dislocation motion and interactions as well as quenched disorder \cite{papanikolaou2012quasi}. By contrast, in our system, lattice distortions are induced by secondary instabilities associated with the air-fluid interface. Further, spatial fluctuations in depth due to micro-scale surface roughness, as well as geometric frustration of the lattice near circular boundaries can serve as sources of quenched disorder in our system. Thus, a detailed theoretical analysis of our results, possibly along the lines of studies on crystal plasticity, would be an exciting topic for future research.

\bibliographystyle{ieeetr}
\bibliography{Faraday.bib}

\vspace{0.5cm}
\subsection*{Acknowledgments} 
We thank John Bush for providing access to his laboratory, where these experiments were conducted.

\subsubsection*{Funding}
V.F. acknowledges the financial support of the National Science Foundation grant No. CMMI-2154151.
S.G. acknowledges the Gordon and Betty Moore Foundation for support as a Physics of Living Systems Fellow through Grant No. GBMF4513.

\subsubsection*{Competing interests}
The authors declare no competing interests.

\subsubsection*{Data and materials availability}
All data are available in the main text, supplemental materials, or from the authors upon request. All code is available from the authors upon request.

\section*{Supplemental Information}

\subsection*{Supplemental videos}
Please email Valeri Frumkin (valerafr@mit.edu) or Shreyas Gokhale (gokhales@mit.edu) for Video S1 and Video S2.

\subsection*{Theoretical prediction for the spatial structure of vibrational modes of a circular elastic membrane}
The normal modes of a vibratiing circular membrane can be calculated analytically by solving the wave equation in a cylindrical geometry with Dirichlet boundary conditions. The general solution for the height profile of normal mode $(m,n)$ is given by 
\begin{equation*}
\begin{split}
     h_{mn}(r,\theta,t) = (A\cos c\lambda_{mn}t + B\sin c\lambda_{mn}t) \times \\
    J_{m}(\lambda_{mn}r)(C\cos m\theta +D\sin m\theta)
\end{split}
\end{equation*}
where $\lambda_{mn} = \alpha_{mn}/R$, $R$ being the membrane radius, and $\alpha_{mn}$ being the $n^{\textrm{th}}$ positive root of the Bessel function of the first kind $J_m$. We set the wave speed $c$ to unity in our simulations. Further, we set the constants $A = C = 0$, and $B = D = 1$, so that $h_{mn}(r,0,t) = h_{mn}(r,\theta,0) = 0$. In our experiments, the pattern intensity is proportional to the magnitude of the slope of the height function, rather than the height itself. To facilitate comparison between theory and experiment, therefore, we define the the predicted intensity profile to be 
\begin{equation*}
I_{mn}(r,\theta,t) = \frac{I_0}{1 + |\nabla h_{mn}(r,\theta,t)|}
\end{equation*}
To ensure clear visualization of the mode structure, in Fig. \ref{fig2}c,d, we set the scale factor $I_0 = 5$. To construct the images shown in Fig. \ref{fig2}c,d, we computed the standard deviation of $I_{mn}(r,\theta,t)$ over one oscillation period, in accordance with the protocol followed for the experimental data. Finally, we manually rotated the obtained intensity patterns to match the global orientation of the corresponding experimental ones. 

\subsection*{Quantification of timescales of quasiperiodic dynamics}
To quantify $F_{10}(t)$, we processed our videos using the software ImageJ. We first analyzed a square region $\approx 10.5\textrm{cm} \times 10.5\textrm{cm}$ in size, in the center of the field of view to minimize the effect of the localized disorder that persists near the circular boundary due to geometric frustration. We then applied the inbuilt ImageJ filter `Convolve', subtracted a constant intensity value, and filtered out noisy pixels using the operation `Despeckle'. Next, we identified regional intensity maxima using the MATLAB function `imregionalmax', and quantified the fraction $F_{10}$ of all pixels associated with regional maxima that are associated with maxima of size exceeding 10 pixels. Finally, we smoothed the trajectories $F_{10}(t)$ in time using successive application of moving mean and moving median filters. To minimize the detection of spurious peaks in $F_{10}(t)$, we only considered peaks with prominence $> 0.002$, and peak separation greater than $10$ s.

\begin{figure*}
  \centering
  \includegraphics[width=1 \textwidth]{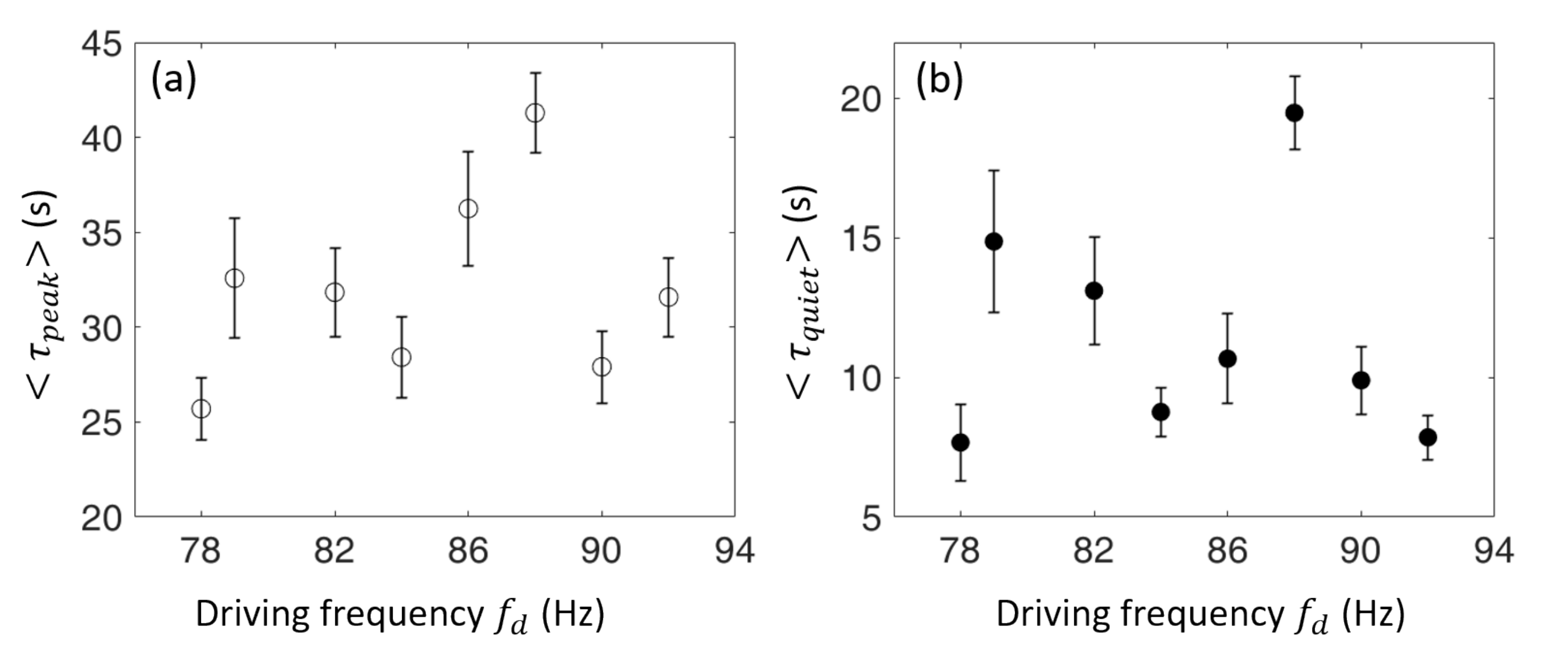}
  \renewcommand{\figurename}
  \caption{{FIG. S1. \bf Timescales of quasiperiodic dynamics for different smoothing parameters.} (a-b) The average distance between successive peaks in $F_{10}(t)$, $\langle \tau_{\textrm{peak}} \rangle$ (a), and the average duration over which the system remains completely ordered, $\langle \tau_{\textrm{quiet}} \rangle$ (b), as a function of driving frequency $f_d$. Error bars are standard errors of the mean of the distributions. To compute the timescales, $F_{10}(t)$ is smoothed in time using a moving mean filter with a window of 10 frames ($0.5$ s), followed by the application of a moving median filters of window width 100 frames ($5$ s), and a second moving median filter of width 20 frames ($1$ s). Clear maxima at $f_d = 88$ Hz and $f_d = 79$ Hz still visible, showing that the qualitative trends are robust to changes in smoothing parameters.}
  \label{figS1}
\end{figure*}

\begin{figure}
  \centering
  \includegraphics[width=0.5 \textwidth]{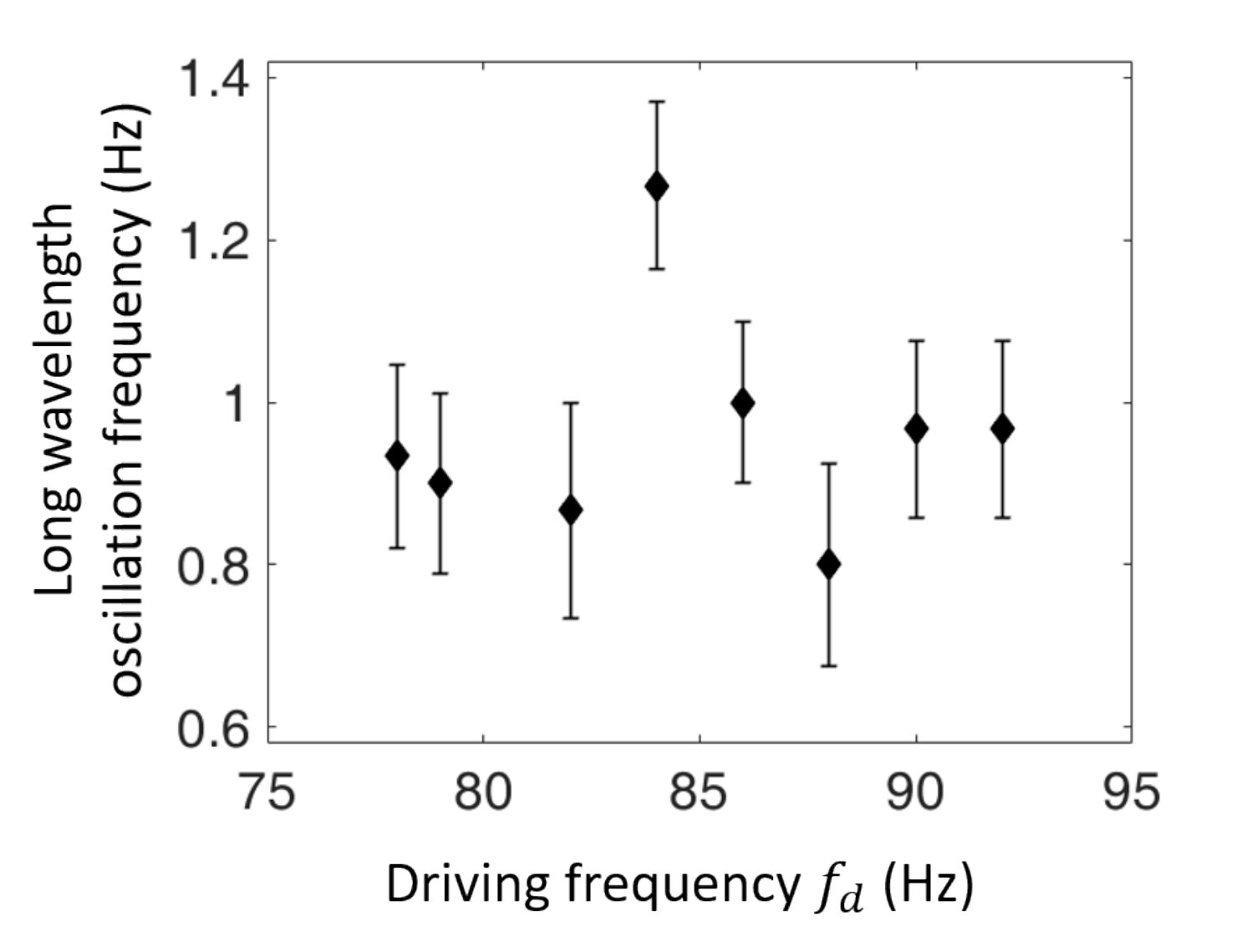}
  \renewcommand{\figurename}
  \caption{FIG. S2. {\bf Frequency of long wavelength modes as a function of driving frequency $f_d$.} Error bars denote combined error due to finite time resolution of the discrete Fourier Transform ($0.1$ Hz), and the standard error of the mean across three distinct quasiperiods.}
  \label{figS2}
\end{figure}

\end{document}